\documentclass[12pt]{article}

\usepackage{graphicx}

\begin{document}

\begin{center}
\vspace{24pt}
{\large \bf A second-order phase transition in CDT}
\vspace{30pt}

{\sl J. Ambj\o rn}$\,^{a}$,
{\sl S. Jordan}$\,^{b}$,
{\sl J. Jurkiewicz}$\,^{c}$
and {\sl R. Loll}$\,^{b}$

\vspace{24pt}

{\footnotesize

$^a$~The Niels Bohr Institute, Copenhagen University,\\
Blegdamsvej 17, DK-2100 Copenhagen \O , Denmark.\\
{email: ambjorn@nbi.dk}\\

\vspace{10pt}

$^b$~Institute for Theoretical Physics, Utrecht University, \\
Leuvenlaan 4, NL-3584 CE Utrecht, The Netherlands.\\
{email: s.jordan@uu.nl, r.loll@uu.nl}\\

\vspace{10pt}

$^c$~Institute of Physics, Jagellonian University,\\
Reymonta 4, PL 30-059 Krakow, Poland.\\
{ email: jurkiewicz@th.if.uj.edu.pl}\\

}

\vspace{48pt}

\end{center}

\begin{center}
{\bf Abstract}
\end{center}
Causal Dynamical Triangulations (CDT) are
a concrete attempt to define a nonperturbative 
path integral for quantum gravity. We present strong evidence that the 
lattice theory has a second-order phase transition line,
which can potentially be used to define a continuum limit
in the conventional sense of nongravitational lattice theories.


\newpage

\section{Introduction}

{\it Dynamical triangulations} (DT) were invented as a nonperturbative
regularization of bosonic string theory and thus also of two-dimensional 
quantum gravity coupled to conformal matter.
This program was both a failure --
in showing that even in a nonperturbative 
setting no bosonic string theory exists in dimension two or larger -- 
and an amazing success, in providing a versatile
regularization of 2d quantum gravity coupled to conformal
matter with central charge $c \leq 1$, i.e. noncritical string theory.
Surprisingly, in many ways the regularized theory turned out to be easier 
to solve analytically than the corresponding continuum theory. 

Encouraged by this, DT was generalized to provide a regularization of 
quantum gravity in three \cite{av,am3,andre}
and four dimensions \cite{aj,am4}. The na\"ive 
expectation was that {\it if} a 
`stand-alone' four-dimensional theory of quantum gravity existed,
the regularized theory should have a second-order phase transition, 
which could be used to define a continuum theory 
of quantum gravity. Second-order transitions are usually characterized
by a divergent correlation length associated with propagating field degree(s) of 
freedom, in the case at hand presumably of a gravitonic nature. 
If a given phase transition point was a UV fixed point, one 
could also attempt to make contact with Weinberg's asymptotic-safety scenario,
for which plenty of corroborating evidence has been found recently 
(see \cite{reuter-review} for reviews). 

A phase transition point was indeed located in 4d DT, and at first 
believed to be of second order \cite{aj,fractal4d,more4d}.
However, analyzing larger lattice systems changed the verdict to a 
first-order phase transition where no obvious continuum limit
could be defined, at least not when using the Regge version of the 
Einstein-Hilbert action \cite{Bialas:1996wu,deBakker:1996zx}. 
Neither did one find convincing 
evidence of a good classical behaviour of large-scale geometry away from the
phase transition.

Partly triggered by this impasse, a modified lattice model 
in terms of {\it Causal Dynamical Triangulations} (CDT) was proposed, and
subsequently shown to have
long-distance properties in agreement with (semi-)classical gravity \cite{cdt}. 
It still uses the Regge-Einstein-Hilbert action, but 
assumes the existence of a global time-foliation, and
has a more complicated phase diagram than the simplest DT 
model.\footnote{The notable similarities with `Ho\v rava-Lifshitz gravity'
\cite{horava} are being explored, see e.g. \cite{Ambjorn:2010hu}.}
In what follows, we will provide strong, new
evidence that -- unlike its Euclidean counterpart -- 4d CDT quantum gravity 
possesses a second-order phase transition line.      

\section{Causal Dynamical Triangulations}

We begin with a brief account of CDT, focusing on several important aspects 
(see 
\cite{ Ambjorn:2005qt,Ambjorn:2001cv,Ambjorn:2007jv,Ambjorn:2008wc}
for technical details and \cite{reviews} for reviews). 
CDT can be characterized as a nonperturbative path integral which 
is as close as possible to a canonical quantization: spacetime histories
share a global foliation, where each leaf is a spatial hypersurface, given in terms of 
a three-dimensional triangulation of fixed topology $\mathcal{T}$, 
built from equilateral 
tetrahedra with link length $a_s$, and labeled by a discrete proper time $t_n$. 
Adjacent hypersurfaces are connected by four-simplices, resulting in 
spacetime histories of the form of four-dimensional 
triangulations of topology $\mathcal{T}\times[0,1]$. 
We use $\mathcal{T}=S^3$ and impose 
periodic boundary conditions in time, such that the spacetime 
topology is $S^3\times S^1$. 
The geometry of each spacetime is fixed by how the four-simplices 
are `glued together' to
form a simplicial manifold, and by the lengths of its links, which come in two types:
{\it spacelike} links which lie entirely within a given hypersurface, and {\it timelike}
links whose endpoints lie on adjacent hypersurfaces, and which have a
(squared) edge length $a_t^2=\alpha a_s^2$, for some fixed relative scaling 
parameter $\alpha<0$. 

The CDT gravitational path integral is a sum over all geometrically inequivalent 
triangulations of this type with a fixed number of time steps, 
with amplitudes depending on the
above-mentioned Regge-Einstein-Hilbert action \cite{Regge:1961px}. 
Since in four dimensions analytical methods are mostly unavailable
we use Monte Carlo simulations to extract physical results. 
To do this we must convert the path integral to a 
statistical partition function by applying a Wick rotation, which 
due to the foliated structure exists globally \cite{Ambjorn:2001cv}. 
It can be implemented  
by rotating 
$\alpha\rightarrow -\alpha$ in the lower-half complex plane, leading to 
the {\it Euclidean} Regge action
\begin{eqnarray}
S_E =\frac{1}{G}\int\sqrt g(-R+2\Lambda) 
\rightarrow -(\kappa_0+6\Delta)N_0+\kappa_4 N_4+
\Delta(N_4+N_4^{(4,1)}) \label{reggeaction},
\end{eqnarray}
where $N_0$ and $N_4$ denote the total number of vertices and four-simplices and 
$N_4^{(4,1)}$ counts the subset of four-simplices which have four vertices on one 
hypersurface and the fifth one on a neighbouring one. 
The couplings $\kappa_0$, $\kappa_4$ and $\Delta$ are 
linearly related to the bare inverse gravitational coupling, 
the bare cosmological coupling and the parameter $\alpha$ 
introduced above. Redefining $\tilde{\kappa}_4=\kappa_4+\Delta$, 
we obtain the Euclidean Regge action
implemented in the computer simulations, namely,
\begin{equation}
S_{Regge} = -(\kappa_0+6\Delta)N_0+
\tilde{\kappa}_4 N_4+\Delta N_4^{(4,1)}=:
-\kappa_0 N_0 +\tilde{\kappa}_4 N_4+\Delta\, \mathrm{conj}(\Delta),
 \label{softwareaction}
\end{equation}
where for later convenience we have introduced the quantity
$\mathrm{conj}(\Delta)=N_4^{(4,1)}-6 N_0$ 
conjugate to $\Delta$. This turns the gravitational
path integral into a statistical 
partition function with Boltzmann weights $\exp (-S_{Regge})$.
We will use the freedom to switch to a different ensemble, obtained
by keeping $N_4$ (measuring the four-volume of the system) fixed, instead of its
conjugate $\kappa_4$. 
In this way we can treat $N_4$ as a finite-size scaling parameter 
which does not appear in the phase 
diagram of the putative continuum theory. The remaining  
couplings $\kappa_0$ and $\Delta$ span the phase diagram, 
which we will go on to explore in the next section.

\section{The phase diagram of CDT}
\label{section:phasediagram}

\begin{figure}[t]
\centerline{\scalebox{1.3}{\includegraphics{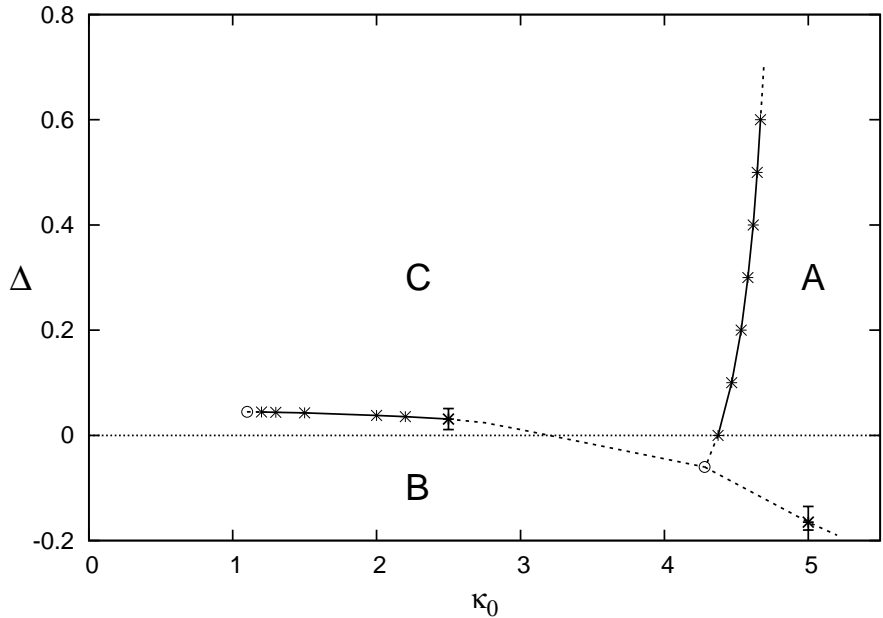}}}
\caption{The phase diagram of CDT. The large crosses represent actual 
measurements.}
\label{fig:pd_plot}
\end{figure}

A qualitative description of the CDT phase diagram first appeared  
in \cite{Ambjorn:2005qt}, with a quantitative plot presented later in 
\cite{Ambjorn:2010hu}. 
The diagram exhibits three phases,
labeled A, B and C in Fig.\ \ref{fig:pd_plot}.
A geometric characterization of the phases can be given in terms of
their distinct spatial volume profiles $N_3(t)$, measuring 
the three-volume in lattice units as a function of proper time $t$. 
As described in detail elsewhere \cite{Ambjorn:2008wc,Ambjorn:2007jv}, 
the average large-scale geometry found in phase C shows 
the scaling behaviour of a genuine four-dimensional universe.
The average volume profile matches beautifully that of a Euclidean 
de Sitter spacetime. Even the quantum fluctuations
around this {\it emergent background geometry} are described well by a 
cosmological minisuperspace action \cite{Ambjorn:2008wc,semi}. 
The situation in the other phases is completely different. 
The typical volume profile 
of a configuration in phase A shows an almost uncorrelated sequence of 
spatial slices, while the configurations in phase B are characterized 
by an almost vanishing time extension. 
 
A preliminary analysis in \cite{Ambjorn:2010hu} suggested that the 
A-C transition is of first order, similar in nature
to the first-order transition observed in the DT formalism mentioned
above. However, as also pointed out in \cite{agjlgt}, to nail down this result 
the numerical evidence still needs to be strengthened.     
By contrast, our interest in the present letter is in the order of the B-C transition, 
about which considerable doubt remained in \cite{Ambjorn:2010hu}. 
We will present strong evidence below that it is a second-order transition line. 

Before doing so, it is instructive to analyze how the transitions change as we 
move along the respective phase transition lines, while holding the system size 
fixed. The A-C transition line is characterized by a jump in
$N_0$, the variable conjugate to the coupling constant $\kappa_0$,
when we cross the line by changing $\kappa_0$. We see no  
appreciable change in this signal when we move along the A-C 
transition line, although we have not examined closely the triple point 
where all three phases meet. 

The situation is very different for the 
B-C transition. When changing the coupling constant $\Delta$ and 
crossing the transition line, we observe a jump in the variable  
$\mathrm{conj}(\Delta)$.
However, moving to smaller values of $\kappa_0$ on the left, the jump 
decreases. Around $\kappa_0=1.0$, no signature of a 
phase transition remains. We conclude that the 
B-C transition has an endpoint, which for $N_4=80k$ is located 
around $\kappa_0=1.0$. Moving to the right the jump increases, and
around $\kappa_0 = 2.3$ the transition is so strong that we 
get stuck in metastable states. The dashed part of the B-C line 
in Fig.\ \ref{fig:pd_plot} marks the region where 
conventional methods are 
insufficient to measure the location of the phase transition with acceptable 
accuracy. We are analyzing currently whether
the use of multicanonical Monte Carlo simulations 
can help in resolving this issue.

\section{The order of the B-C phase transition}
\label{sec:ptorder}

Measuring the order of phase transitions requires some care. 
To confirm that a phase transition is {\it not} a first-, but a 
second-order transition, one can try to measure 
various so-called critical exponents.  
One such exponent measures the shift of a transition point with system size. 
Recall first how this works for a conventional system such as the 
Ising model with volume $V=L^d$, where $d$ is the dimension of the 
system \cite{Newman:1999a}. Considering the Ising model's
temperature-driven phase 
transition and using the location of the maximum of 
the magnetic susceptibility to define a transition point $\beta^c(V)$, 
one finds the power-law behaviour
\begin{equation}
|\beta^c(\infty)-\beta^c(V)|\propto V^{-1/\nu d}
\label{eq:ptshift}
\end{equation}
for sufficiently large system sizes.
The exponent $\nu$ governs the increase of the correlation length in 
a second-order transition as one moves towards the 
critical point $\beta^c(\infty)$ on an infinite lattice. 
For first-order transitions 
there is no correlation length and one expects a 
scaling like (\ref{eq:ptshift}), with $\nu d$ replaced by 
an exponent $\tilde{\nu}$ where $\tilde{\nu}=1$ \cite{MeyerOrtmanns:1996ea}. 
A sufficiently strong violation of $\tilde{\nu}=1$ therefore
signals the presence of a second-order transition.
Another quantity of interest is the Binder cumulant 
\begin{equation}
B_{\cal O}=\frac{1}{3}\left(1-\frac{\left<{\cal O}^4\right>}
{\left<{\cal O}^2\right>^2}\right)
\label{binder}
\end{equation}
associated with an observable $\cal O$ \cite{MeyerOrtmanns:1996ea},
which is always nonpositive and zero if the histogram of 
$\cal O$ is Gaussian. Evaluating $B_{\cal O}$ as a function of the couplings, 
its local minima lie at transition points. We can measure these 
minima for different system sizes and by extrapolation determine 
$B_{\cal O}^{\min}(1/N_4=0)$. At a second-order transition the 
histogram of $\cal O$ should converge to a single Gaussian distribution
with the Binder cumulant going to zero. At a first-order transition it 
will go to a nonzero constant, however, its value at a weak first-order
transition can be small.
 
\begin{figure}[t]
\centerline{\scalebox{1.3}{\includegraphics{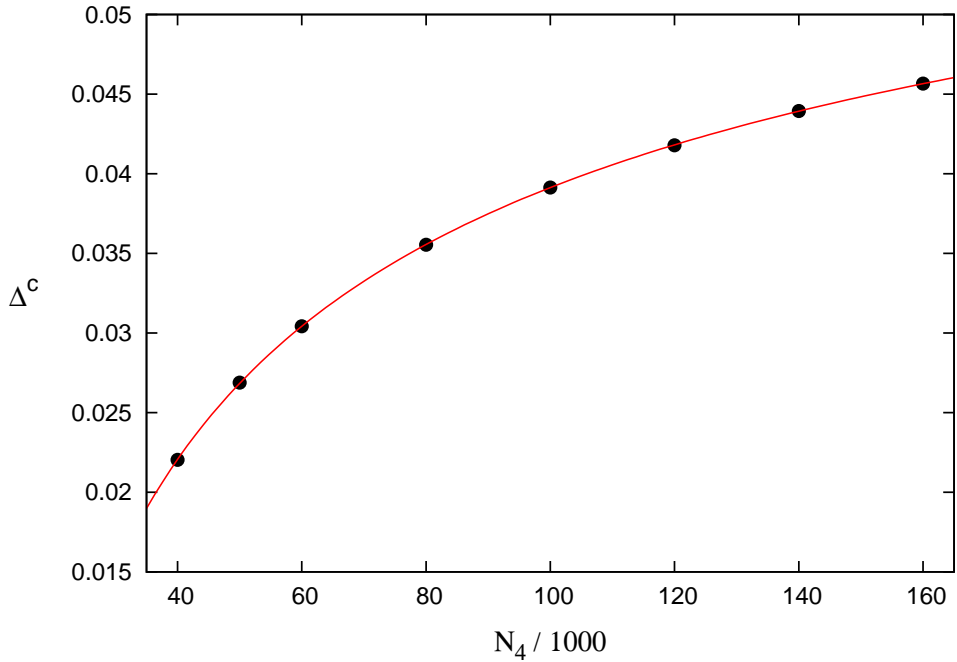}}}
\caption{Measuring the location $\Delta^c$ of B-C transition points at 
$\kappa_0=2.2$ for different system sizes $N_4$ 
to determine the shift exponent $\tilde{\nu}$.}
\label{fig:shiftexpBC}
\end{figure}

To analyze the B-C transition, we fixed $\kappa_0=2.2$  and used systems 
of size 40, 50, 60, 80, 100, 120, 140 and 160$k$. The number of 
sweeps used was approximately $2.5\cdot 10^6$, with
one sweep consisting of one million attempted Monte Carlo moves.  
We have measured the shift exponent $\tilde\nu$ for the asymmetry
parameter $\Delta$ using 
\begin{equation}
\Delta^c(N_4)=\Delta^c(\infty)-C N_4^{-1/\tilde{\nu}},
\label{eq:ptshiftBC}
\end{equation}
where $C$ is a proportionality factor, and $\Delta^c$ has been defined
using the location of the maximum of the susceptibility 
$\chi_{\mathrm{conj}(\Delta)}\! =\!
\left<\mathrm{conj}(\Delta)^2\right>\! -\!\left<\mathrm{conj}(\Delta)\right>^2$. 
Fig.\ \ref{fig:shiftexpBC} shows the measured data points (error bars 
too small to be included) and the best fit through all of them, 
yielding $\tilde{\nu}=2.39(3)$. To judge whether 
our range of system sizes lies inside the scaling region we have
performed a sequence of fits by successively removing the data points 
of lowest four-volume, leading to $\tilde\nu$-values 
$2.51(3)$, $2.49(3)$ and
$2.51(5)$, where the last fit was performed with all but five data 
points removed. 
This suggests that the data point with the lowest four-volume lies 
outside the scaling region. Removing it from the fit we obtain
\begin{equation}
\tilde{\nu}=2.51(3).
\end{equation}
This result makes a strong case for a second-order transition, 
since the prediction $\tilde{\nu}=1$ for a first-order transition is 
clearly violated. (By contrast, for the A-C transition one finds 
$\tilde{\nu} \approx 1$
as will be reported elsewhere \cite{to-appear}.) 

\begin{figure}[t]
\centerline{\scalebox{1.3}{\includegraphics{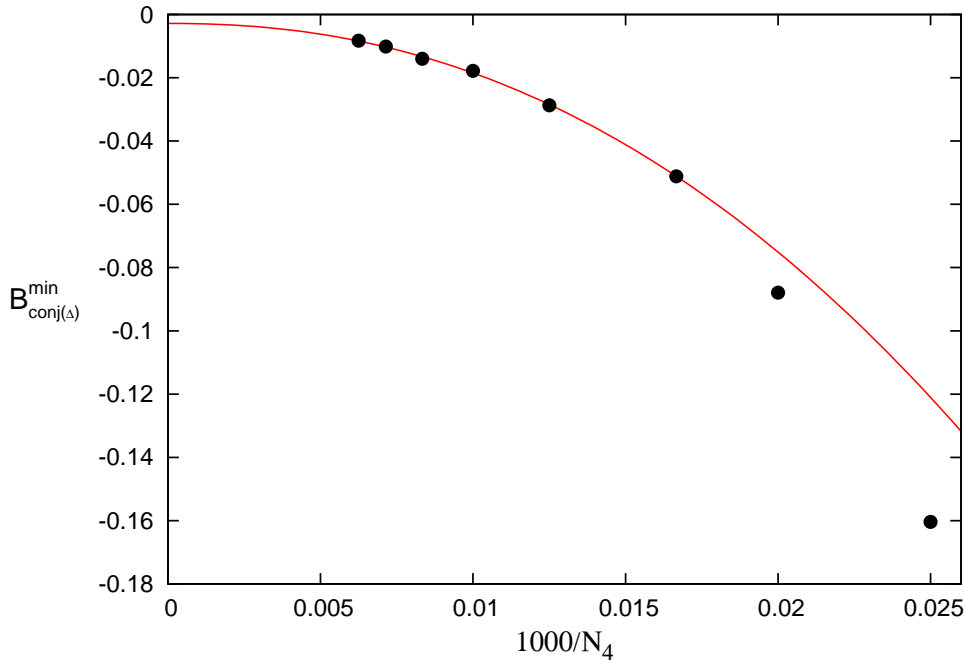}}}
\caption{Dependence of the minimum of the 
Binder cumulant $B_{\mathrm{conj}(\Delta)}$ on the (inverse)
system size at the B-C 
transition. At a second-order transition, $B^{\min}\!\rightarrow\! 0$ in
the infinite-volume limit. (Fit excludes the two points on the right.)}
\label{fig:bincumBC}
\end{figure}

Lastly, we have investigated how the minimum of the Binder cumulant 
(\ref{binder}) depends on the system size. Fig.\ \ref{fig:bincumBC} 
shows $B_{\mathrm{conj}(\Delta)}^{\min}$ as a function of
\emph{inverse} system size (errors
approximately equal to the dot radii). 
Inside the scaling region the minimum of the cumulant is expected 
to have a power-law behaviour. To understand which data 
points lie inside the scaling region, we have again performed
a sequence of fits by successively removing the points 
of lowest four-volume. This has led to the exclusion of 
the data points for $N_4=40k$ and 
$N_4=50k$ from the fit shown in Fig.\ \ref{fig:bincumBC}. 

\begin{table}[t]
\begin{center}
\begin{tabular}{|c|r|}
\hline
observable $\cal O$ & $B_{\cal O}^{\min}(N_4\rightarrow\infty)$ \\
\hline
\hline
$\mathrm{conj}(\Delta)$ & $-0.003(4)$ \\
\hline
$N_4^{(4,1)}$ & $-0.001(3)$ \\
\hline
$N_2$ & $-0.0000001(3)$ \\
\hline
$N_1$ & $-0.000003(7)$ \\
\hline
$N_0$ & $0.0000(3)$ \\
\hline
\end{tabular}
\end{center}
\caption{Measurements of $B_{\cal O}^{\min}(N_4\!\rightarrow\!\infty)$ 
for various observables $\cal O$, where $N_k$ denotes the number 
of $k$-dimensional (sub-)simplices in the triangulation.}
\label{tab:bincumBC}
\end{table}

Table \ref{tab:bincumBC} collects the extrapolations 
$B_{\cal O}^{\min}(N_4\!\rightarrow\!\infty)$ for several observables $\cal O$. 
As indicated earlier, it is always difficult to make a strong case 
for a second-order transition based on Binder cumulant measurements alone, 
because weak first-order transitions may show a convergence to a 
nonzero value close to zero. 
Nevertheless, in the case at hand all our measurements are mutually
consistent and in excellent agreement with the limiting value 0,
further corroborating our claim of the second-order nature of the transition.

\section{Discussion}

We have succeeded in our goal of determining the order of the B-C 
transition in CDT quantum gravity by applying two distinct methods, namely, 
measuring the shift exponent and analyzing Binder cumulants. 
The measured shift exponent 
$\tilde{\nu}=2.51(3)$ represents a strong violation of the 
prediction $\tilde{\nu}=1$ for a first-order transition. 
Also the results of the Binder cumulant 
analysis are clearly and unambiguously consistent with the 
second-order nature of the transition.

From this we conclude that there is strong evidence 
that the B-C transition is of second order, making four-dimensional
CDT quantum gravity the first known instance of a dynamically triangulated
model (without matter coupling) in any signature and dimension which
displays such a transition. This result is
potentially very attractive. It opens the door to studying
critical phenomena in CDT and 
defining a continuum limit where the lattice spacing (the UV cut-off) 
is taken to zero, just as one does in standard lattice quantum field theories
with nondynamical geometry.

\vspace{.5cm}

\noindent {\bf Acknowledgements.} SJ thanks Prof.\ G.T.\ Barkema 
for fruitful discussions and valuable advice on 
the numerical aspects of this work. JA thanks the ITP, Utrecht for hospitality.
RL acknowledges support by the Netherlands
Organisation for Scientific Research (NWO) under their VICI
program. The contributions by SJ and RL are 
part of the research programme of the 
Foundation for Fundamental Research on Matter (FOM), financially 
supported by NWO.


\end{document}